\documentclass[preprint,12pt]{elsarticle}

\usepackage{amssymb}
\usepackage{amsmath}
\usepackage{optidef}
\usepackage{xcolor}
\usepackage{siunitx}
\usepackage{multirow}
\usepackage{graphicx}
\usepackage{subcaption}

\usepackage{tikz}								
\usepackage[american, cuteinductors,smartlabels, fulldiode, siunitx, americanvoltages, oldvoltagedirection, smartlabels]{circuitikz}
\tikzstyle{block} = [draw, rectangle, 
    minimum height=2em, minimum width=4em]
    \tikzstyle{block2} = [draw, rectangle, 
    minimum height=2em, minimum width=2em]
\tikzstyle{sum} = [draw, circle, node distance=1cm]
\tikzstyle{input} = [coordinate]
\tikzstyle{output} = [coordinate]
\tikzstyle{tmp} = [coordinate]

\usepackage[ruled,vlined,linesnumbered,noresetcount]{algorithm2e}

\begin{document}

\newcommand{\norm}[1]{|| #1 ||}
\newcommand{\Hone}{\mathcal{H}_{1}}
\newcommand{\Htwo}{\mathcal{H}_{2}}
\newcommand{\Hinf}{\mathcal{H}_{\infty}}
\newcommand{\argmin}[1]{argmin( #1 )}

\begin{frontmatter}



\title{Virtual Reference Feedback Tuning with robustness constraints: a swarm intelligence solution}

\author[inst0,inst1]{Luan Vinícius Fiorio\corref{cor1}}
\ead{l.v.fiorio@tue.nl}
\cortext[cor1]{Corresponding author.}
\author[inst2,inst3]{Chrystian Lenon Remes}
\author[inst4]{Patrick Wheeler}
\author[inst1,inst4]{Yales Rômulo de Novaes}

\affiliation[inst0]{department={Department of Electrical Engineering,},
            organization={Eindhoven University of Technology},
            postcode={5600 MB}, 
            city={Eindhoven},            
            country={the Netherlands}}

\affiliation[inst1]{department={Department of Electrical Engineering,},
            organization={Santa Catarina State University},
            postcode={89219-710},
            city={Joinville},            
            country={Brazil}}

\affiliation[inst2]{department={Department of Electrical Engineering,},
            organization={Federal University of Paraná},
            postcode={80060-000}, 
            city={Curitiba},
            country={Brazil}}
            
\affiliation[inst3]{department={Department of Electrical Engineering,},
            organization={Federal University of Rio Grande do Sul},
            postcode={90035-190},
            city={Porto Alegre},            
            country={Brazil}}

\affiliation[inst4]{department={Department of Electrical and Electronic Engineering,},
            organization={University of Nottingham},
            postcode={NG7 2RD}, 
            city={Nottingham},            
            country={United Kingdom}}

\begin{abstract}
The simplified modeling of a complex system allied with a low-order controller structure can lead to poor closed-loop performance and robustness. A feasible solution is to avoid the necessity of a model by using data for the controller design. The Virtual Reference Feedback Tuning (VRFT) is a data-driven design method that only requires a single batch of data and solves a reference tracking problem, although with no guarantee of robustness. In this work, the inclusion of an $\Hinf$ robustness constraint to the VRFT cost function is addressed. The estimation of the $\Hinf$ norm of the sensitivity transfer function is extended to maintain the one-shot characteristic of the VRFT. Swarm intelligence algorithms are used to solve the non-convex cost function. The proposed method is applied in two real-world inspired problems with four different swarm intelligence algorithms, which are compared with each other through a Monte Carlo experiment of 50 executions. The obtained results are satisfactory, achieving the desired robustness criteria.

\end{abstract}


\end{frontmatter}


\section{Introduction}
\label{sec:introduction}
The complexity of certain processes usually requires simplification in the mathematical modeling to allow proper controller design \cite{Nise2000}, e.g., for power systems \cite{Chaudhuri2012,Xie2021}. For dc-dc converters, for example, the majority of the control techniques assume the existence of an accurate model \cite{Kazimierczuk2008,Kobaku2017}, which can be difficult to obtain since power converters have nonlinear dynamics. Low-order controllers are commonly used in the industry, such as Proportional Integral (PI) and Proportional Integral Derivative (PID) controllers \cite{Aguiar2018,Tharanidharan2022,Tudon-Martinez2022,vanTan20222}, but a robust design can still be a challenge. Oversimplified modeling due to the process complexity, and/or a limited performance of the chosen controller structure are possible reasons for achieving poor performance and robustness of the controlled system \cite{Keel2008}.

On the other hand, data-driven control design techniques are used to overcome common problems related to system models, such as the dilemma on representativity and complexity, or even the unavailability of those \cite{Remes2021,Zenelis2022,Huang2022}. Some of the data-driven approaches require several plant experiments and iterative acquisition of data, like Iterative Feedback Tuning (IFT) \cite{Hjalmarsson1998}, Correlation-based Tuning (CbT) \cite{Karimi2004}, and more recent methods such as the deep neural network-based implementation of data-driven Iterative Learning Control (ILC) \cite{Li2020}, and data-driven optimization-based ILC \cite{Meng2022}. While other methods like Virtual Reference Feedback Tuning (VRFT) \cite{Campi2002}, Direct Iterative Tuning (DIT) \cite{Kammer2000}, Optimal Controller Identification (OCI) \cite{Campestrini2017}, Virtual Disturbance Feedback Tuning (VDFT) \cite{Eckhard2018}, Data-Driven Linear Quadratic Regulator (DD-LQR) \cite{Goncalves2019}, and data-driven predictive model control \cite{Carlet2020} only require a single batch of data (one-shot) in order to tune the controller parameters. If data acquisition is difficult, its repetition is prevented by a tight deadline, or memory constraints are an issue, one-shot solutions are usually preferred over iterative solutions. For more details, a recent and systematic review of data-driven control methods can be found at \cite{Prag2022}.

Robustness considering low-order controllers is a frequent topic of discussion \cite{Perez2018,Alcantara2013} since specific processes can present uncertainties, as well as disturbances that might occur over time. Another point to be observed is that a poor choice of reference model or limited controller class in, e.g., the VRFT design, may result in poor performance or robustness \cite{Bazanella2014}. Since robustness can be measured by the $\Hinf$ norm of the sensitivity transfer function $S(z)$ of a closed-loop system \cite{Skogestad2005}, its inclusion in the data-driven design of controllers can be considered, allowing for a more robust design when necessary. A recent methodology proposed in \cite{Chiluka2021} has suggested the inclusion of a robustness criterion in the VRFT design, at the expense of: i) more experiments since the proposed design procedure iterates in a trial-and-error fashion until the desired robustness is achieved and, essentially removes one of the greatest advantages of the VRFT - being a one-shot method; and ii) this type of iterative procedure requiring more background knowledge from the designer for choosing reference models and specifying requirements.

Also, a data-driven one-shot approach for multivariable systems regarding robust solutions of $\Htwo$ and $\Hinf$ criteria, and loop-shaping specifications, has been recently addressed \cite{Karimi2017}. But the method relies on an initial solution that influences the final one since it is convexified by linearization around the initial stabilizing controller, which is undesirable. In the case of data-driven one-shot approaches for Single-Input Single-Output (SISO) systems, considering an $\Hinf$ robust performance criteria, a recent work \cite{Nicoletti2019} proposes a technique where the order of the controller is increased as the solution converges and only deals with noise-free data. Both methods \cite{Karimi2017, Nicoletti2019} use only frequency-domain data. Other data-driven robust solutions are achieved in an online fashion and require higher computational processing than offline techniques, and also present the demand of measuring signals in real-time. Some of those techniques are the use of the modified Riccati equation with online data-driven learning \cite{Na2021} and the application of a data-driven Model Predictive Control method with robustness guarantees \cite{Berberich2021}.

Among the aforementioned data-driven design techniques, the VRFT has been more broadly applied to several classes of problems, a fact that can be useful to attest to the feasibility of the proposed method of this paper, as it requires less computational cost than, e.g., OCI and online methods. Therefore, this work is based on the VRFT technique, as presented in \cite{Campi2002} and \cite{Campestrini2011}, proposing the inclusion of a robustness constraint, in terms of the $\Hinf$ norm of the sensitivity transfer function of the system, in the VRFT design, but maintaining one of its most attractive features: the necessity of only one batch of data. The constraint is included in the VRFT cost function as a penalty \cite{Luenberger2015}, which compromises its convex behavior. To deal with the non-convex optimization procedure, the proposed method is tackled in two main steps:
i) the design, in a data-driven fashion, of a controller using the VRFT approach, if a previous controller is unexistent; and ii) the application of a metaheuristic optimization algorithm to minimize the cost function considering an $\norm{S(z)}_{\infty}$ norm constraint, using the controller from the previous step as an initial solution.

Notice that metaheuristics, more specifically the Particle Swarm Optimization (PSO) algorithm, have been used on the VRFT design for enhancing the performance of the controller in terms of reference tracking on hybrid systems \cite{Jiaqi2017}. Also, more closely related to this work, PSO has been used for solving the VRFT with a modified cost function which includes the optimization of the reference model based on a reference tracking metric \cite{Selvi2018}. The work focuses on reducing the mismatch between the reference model and the actual controlled system but does not mention robustness.

Since metaheuristic algorithms may work well for a certain class of problems, but might fail over other problems, according to the No Free Lunch (NFL) theorems \cite{Wolpert:NFL:1997}, more than a single metaheuristic optimization algorithm should be evaluated. Looking over the available types of metaheuristics, three can be highlighted: evolutionary algorithms \cite{Mirjalili2019}; physics-based algorithms \cite{Alatas2015}; and swarm intelligence algorithms. Although some authors group evolutionary algorithms with multiple agents and swarm algorithms together \cite{Wahab2015}, this paper considers the two classes separately, as done by other authors in the metaheuristic optimization subject \cite{Mirjalili:GWO:2014}. This work focuses on swarm intelligence algorithms since they usually have fewer parameters to be tuned by the user or designer \cite{Wahab2015}. Four swarm intelligence algorithms are considered: Particle Swarm Optimization \cite{Kennedy:PSO:1995} and Artificial Bee Colony (ABC) \cite{Karabog:ABC:2005} since those are two of the most used in literature; and the two swarm intelligence algorithms with the least number of hyperparameters, Grey Wolf Optimizer (GWO) \cite{Mirjalili:GWO:2014}, and its most recent version, the Improved Grey Wolf Optimizer (I-GWO) \cite{Nadimi:IGWO:2021}.

In summary, the main contributions of this work are:
\begin{itemize}
    \item in theoretical terms, the derivation of the  expressions of the signals for the data-driven estimation of the impulse response of the sensitivity transfer function of a system, in one-shot, at Subsection~\ref{sssec:signals_S};
    \item the proposal of a modified cost function to the VRFT method regarding the inclusion of a robustness constraint, at Section~\ref{sec:method}, and the use of swarm-intelligence algorithms for solving the proposed cost function;
    \item at Section~\ref{sec:examples}, the use and comparison of four different swarm-intelligence algorithms - PSO, ABC, GWO, and I-GWO - in the proposed method for solving two real-world inspired problems based on the structure of widely used DC-DC converters.
\end{itemize}

This paper is structured as follows: Section~\ref{sec:preliminaries} describes the system that is considered in this paper for the theoretical formulation; Section~\ref{sec:vrft} presents the basis of data-driven controller design and details the basic VRFT design procedure; Section~\ref{sec:Ms} explains the method for the $\Hinf$ norm estimation of the sensitivity transfer function; Section~\ref{sec:method} presents the proposed method and details the four considered swarm intelligence algorithms; Section~\ref{sec:examples} illustrates and validates the method by showing its application in two real-world inspired examples; and finally, Section~\ref{sec:conclusion} concludes this work.

\section{Preliminaries: description of the system}
\label{sec:preliminaries}
The system considered in this paper for the theoretical formulation is a discrete-time, causal, linear time-invariant, and Single-Input Single-Output (SISO) system $G(z)$. It is considered that $z$ is the forward discrete time-shift operator such that $zx(k) = x(k+1)$. The output $y(k)$ of the system can be described as
\begin{equation}
\label{c4:eq:y1}
    y(k) = G(z) u(k) + v(k),
\end{equation}
where $u(k)$ is the input signal and $v(k)$ is the process noise - stochastic effects that are not represented by $G(z)$, i.e., not captured by the input-output relation of $u(k)$ and $y(k)$. 

\begin{figure}[!t]
    \centering
        \begin{tikzpicture}[auto, node distance=1.5cm]
            \node [input, name=input] {};
            \node [sum, right of=input] (sum) {};
            \node [block, right of=sum, node distance=2cm] (controller) {$C(z)$};
            \node [block, right of=controller, node distance=3cm] (system) {$G(z)$};
    
            \draw [->] (controller) -- node[name=u] {$u(k)$} (system);
            \node [sum, right of=system, node distance = 2cm] (sum2) {};
            \node [output, right of=sum2, node distance = 2cm] (output) {};
            
            \draw [draw,->] (input) -- node[pos=0.2] {$r(k) \ \ +$} (sum);
            \node [input, name=noise_input, above of=sum2, node distance = 1cm] {};
            \draw [draw,->] (noise_input) node[above of=sum2,node distance=1.3cm] {$v(k)$} -- node[near end] {$+$} (sum2);
            \draw [->] (sum) -- node {$e(k)$} (controller);
            \draw [->] (system) -- node [near end] {$+$} (sum2);
            \draw [->] (sum2) -- node [name=y] {$y(k)$}(output);
            \node [tmp, below of=u] (tmp1){$H(z)$};
            \draw [->] (y) |- (tmp1)-| node[pos=0.99] {$-$} (sum);
        \end{tikzpicture}
    \caption{Block diagram of the considered closed-loop system structure for this paper.}  
    \label{fig:block_casestudy}
\end{figure}
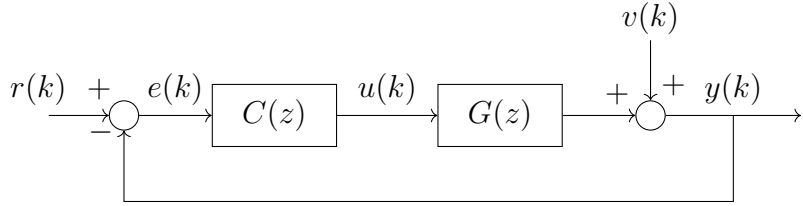

The closed-loop system taken into account in this work is regarded as a controller $C(z)$ with the process $G(z)$ and a unit gain feedback, as shown in Figure~\ref{fig:block_casestudy}, where $r(k)$ is the reference signal and $e(k)$ is the error signal. The closed-loop control law is
\begin{equation}
    u(k) = C(z,\rho) (r(k) - y(k)),
\end{equation}
with 
\begin{equation}
\label{eq:c-rhobar}
    C(z,\rho) = \rho' \bar{C}(z)
\end{equation}
being a controller with parameter $\rho \in \mathbb{R}^p$, $\bar{C}(z)$ a vector of transfer functions belonging to the controller class $\mathcal{C}$ (e.g., PI or PID controller classes) and $r(k)$ the reference signal.

The output of the closed-loop system is given as
\begin{equation}
    y(k) = T(z) r(k) + S(z)v(k),
\end{equation}
where the reference signal $r(k)$ is applied to the transfer function from the reference $r(k)$ to the output $y(k)$, $T(z)$, with
\begin{equation}
\label{eq:T}
    T(z) = \frac{C(z)G(z)}{1 + C(z)G(z)},
\end{equation}
and $S(z)$ is the sensitivity transfer function such that $S(z) + T(z) = 1$ and
\begin{equation}
    \label{eq:sensitivity}
     S(z) = \frac{1}{1 + C(z)G(z)}.
\end{equation}

In the next section, the Model Reference Control (MRC) is introduced. The VRFT method is described for the cases where the process has minimum or non-minimum phase.

\section{Virtual Reference Feedback Tuning (VRFT)}
\label{sec:vrft}

The Model Reference Control (MRC), which is the basis for the VRFT, more generally called model matching control \cite{Goodwin2000}, concerns the problem of reference tracking of the closed-loop system's response, disregarding the effects of noise at the output \cite{Bazanella2014}.

In order to obtain a controller, the MRC requires the designer to elaborate a target transfer function for the controlled closed-loop system, called a reference model ($T_d(z)$), which generates the output $y_d(k) = T_d(z) r(k)$. A reference tracking performance criterion  evaluated by the two-norm tracking error is then obtained by solving the optimization problem
\begin{mini}|l|
  {\rho}{J^{MR}(\rho) = \norm{(T(z,\rho) - T_d(z)) r(k)}_2^2}{}{},
  \label{eq:mrc_J}
\end{mini}
which can be solved by considering \eqref{eq:T}, resulting in the solution controller for the MRC, called the ideal controller $C_d(z)$.

The VRFT is a one-shot optimization data-driven controller design technique based on the MRC. It is defined as one-shot since only a single batch of input-output data is required to solve the model reference control problem \eqref{eq:mrc_J}, which can be done by the use of least squares when the controller is linearly parametrized as in \eqref{eq:c-rhobar}, resulting in the parameter $\rho$ of a controller with predefined class. The VRFT design depicted in this paper follows the procedures of \cite{Bazanella2014,Remes2021}.

\begin{figure}[!t]
    \centering
        \includegraphics[width=0.7\linewidth]{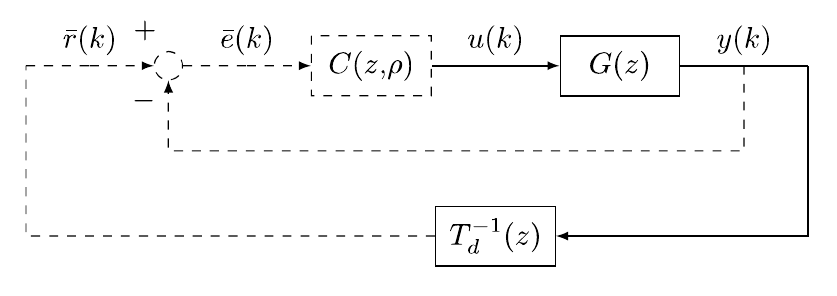}
        \caption{Closed-loop block diagram for the VRFT controller design.}
    \label{fig:vrft} 
\end{figure}

Consider an experiment, in open-loop or closed-loop, that results in a batch of collected data $\{ u,y\}_{k=1}^{N}$. A \textit{virtual reference} signal $\bar{r}(k)$ is defined such that $T_d(z)\bar{r}(k) = y(k)$. A virtual error can be obtained as $\bar{e}(k) = \bar{r}(k) - y(k) = (T_d^{-1}(z) - 1) y(k)$. In summary, a controller $C(z,\rho) = \rho' \bar{C}(z)$ is considered satisfactory if it generates $u(k)$ when fed by $\bar{e}(k)$. The closed-loop of such block diagram for the VRFT controller design is illustrated in Figure~\ref{fig:vrft}.

The VRFT solves the optimization problem
\begin{mini}|l|
    {\rho}{J^{VR}(\rho) = \norm{u(k) - C(z,\rho) (T_d^{-1}(z) - 1) y(k)}_{2}^{2}}{}{},
    \label{eq:vrft_j}      
\end{mini}
having the same minimum as \eqref{eq:mrc_J} if the ideal controller $C_d(z)$ in \eqref{eq:T} belongs to the same controller class $\mathcal{C} = \{ C(z,\rho), \rho \in \mathbb{R}^p \}$ as $C(z,\rho)$. To compensate for the fact that the ideal controller rarely belongs to the chosen controller class, a filter $L(z)$ is applied to the data to approximate the minimum of $J^{VR}$ to the minimum of $J^{MR}$, where the amplitude should satisfy \cite{Bazanella2014}:
\begin{equation}
\label{eq:filter}
    |L(e^{j\Omega})|^2 = |T_d(e^{j\Omega})|^2 |1 - T_d(e^{j\Omega})|^2 \frac{\Phi_r(e^{j\Omega})}{\Phi_u (e^{j\Omega})}, \quad \forall \Omega \in [-\pi,\pi],
\end{equation}
where $x(e^{j\Omega})$, with $x$ representing any signal or system, represents the Discrete Fourier Transform of $x(k)$, $\Phi_r(e^{j\Omega}),\Phi_u (e^{j\Omega})$ are, respectively, the power spectra of the signals $r(k),u(k)$. 

Instrumental variables can be used in order to suppress estimation bias caused by the noise in data \cite{Ljung1999}, requiring the use of a second data batch. In practice, the input signal can be formed by two identical sequences in the same experiment, if memory restrictions do not impose a problem. Then, the signals can be synced together afterward, resulting in two batches of data from one single experiment. 

In the presence of a Non-Minimum Phase (NMP) zero at the process, a flexible reference model can be used, as presented in \cite{Bazanella2014}. The optimization problem \eqref{eq:vrft_j}, then, becomes
\begin{mini}|l|
    {\rho}{J^{VR}(\rho) = \norm{\eta' F(z) (u(k) + \rho' \bar{C}(z) y(k)) - \rho' \bar{C}(z) y(k)}_{2}^{2}}{}{},
    \label{eq:vrft_j_f}  
\end{mini}
where $\eta \in \mathbb{R}^{m}$, and $F(z)$ is a vector of transfer functions such that $T_d(z,\eta) = \eta' F(z)$. The step-by-step design for the VRFT with a flexible reference model, from data collection to the algorithm design, is detailed in \cite{Remes2021}.

To be able to include a robustness criterion at the VRFT cost function, a means of evaluating this parameter is required. Nonetheless, this is approached in the next section.

\section{Robustness index estimation}
\label{sec:Ms}

Depending on the choice of $T_d(z)$ or the controller class $\mathcal{C}$, as well as the response of the plant $G(z)$, the VRFT-designed controller can result in poor robustness of the controlled process. For such cases, a robustness constraint can be included in the form of the $\Hinf$ norm of $S$, also called maximum sensitivity ($M_S$), which can be used as a measure of robustness \cite{Skogestad2005}.

Typically, a system that presents $M_S > 2$ is considered to have poor robustness \cite{Skogestad2005}. In this context, and considering the use of data-driven design approaches, it is necessary to estimate the value of $\norm{S(z)}_{\infty}$ in a data-driven way since it is assumed that no plant model is available to the designer. By this reasoning, the estimation of $M_S$ is explained in the following subsection.

\subsection{Estimation of $M_S$}
\label{ssec:ms_est}

The $\Hinf$ norm estimation procedure developed in this work is based on the Impulse Response (IR) of the system, as presented in \cite{Fiorio2022}, modified from \cite{Goncalves2020} to a SISO impulse response identification procedure, which allows for a regularized estimation according to the existing literature \cite{Chen2012}. Also, in order to maintain the one-shot characteristic of the VRFT, the estimation of the $\Hinf$ norm of $S$ based on impulse response is addressed as follows.

Consider the linear discrete-time causal and SISO system $S$, represented by its transfer function $S(z,\rho)$, such that its output signal $\psi(k)$ is given by the equation
\begin{equation}
		S : \psi(k) = s(k)\ast \zeta(k) = \sum_{n=0}^{\infty}s(k-n)\zeta(n),
		\label{eq:sk}
\end{equation}
where $\zeta(k)$ is the input signal of $S$, whose impulse response is $s(k)$.

Since \eqref{eq:sk} requires infinite data to be obtained, an order $M$ is defined such that it is assumed that any IR term greater than $M$ is negligible, which is valid for stable systems since $\lim_{k\to\infty}s(k) = 0$. Nevertheless, equation \eqref{eq:sk} can be truncated to $M$ terms, leading to:
\begin{equation}
		S : \psi(k) = \sum_{n=0}^{\infty}s(k-n)\zeta(n) \approx \underbrace{\sum_{n=0}^{M}s(k-n)\zeta(n)}_{|s(M+1)|< \epsilon\text{, with }\epsilon\to 0^+}.
		\label{eq:sk_approx}
\end{equation}

On the other hand, the definition of the $\Hinf$ norm, when applied to the system $S$, can be written as \cite{Skogestad2005}
\begin{equation}
\label{eq:sinf_def}
	\Hinf : \norm{S}_{\infty} = \max_{\zeta(k)\neq 0}\dfrac{\norm{s(k)\ast \zeta(k)}_{2}}{\norm{\zeta(k)}_{2}},
\end{equation}
which requires the whole set of possible inputs $\{\zeta(k) \neq 0\}$. Therefore, expression \eqref{eq:sinf_def} cannot be directly calculated. An alternative strategy is to obtain a matrix relation for $S$, which allows for the use of induced norm properties. 

Expanding \eqref{eq:sk_approx} to the $M$ first terms gives
\begin{equation}
	\begin{cases}
		& \psi(0) = s(0) \zeta(0)\\
		& \psi(1) = s(1) \zeta(0) + s(0) \zeta(1)\\
		& \vdots \\
		& \psi(M) = s(M)\zeta(0) + \cdots + s(0) \zeta(M),
	\end{cases}
	\label{eq:S_expanded}
\end{equation}
and the following matrix relation truncated at $M$ elements is obtained:
\begin{equation}
	\underbrace{
	\begin{bmatrix}
		\psi(0) \\ \psi(1) \\ \cdots \\ \psi(M)
	\end{bmatrix}
	}_{\Psi_M}
	 = 
	\underbrace{
	\begin{bmatrix}
		s(0) 		& 0 			& \cdots &		0		\\
		s(1) 		& s(0)		&	\cdots &		0		\\
		\vdots 	& \vdots	& \ddots & \vdots \\
		s(M)		& s(M-1)	& \cdots & 	s(0)
	\end{bmatrix}
	}_{S_M}
	\underbrace{
	\begin{bmatrix}
		\zeta(0) \\ \zeta(1) \\ \cdots \\ \zeta(M)
	\end{bmatrix} 
	}_{Z_M}
	.
	\label{eq:SM}
\end{equation}
From the assumption that the order $M$ is sufficiently high, it can be said that matrix $S_M$ characterizes the IR $s(k)$ of $S$.

A useful matrix property is the induced norm \cite[A.5]{Skogestad2005}, which can be applied to \eqref{eq:SM}, such that
\begin{equation}
		\norm{S_M}_{ip} = \max_{Z_M\neq 0}\dfrac{\norm{S_M Z_M}_{p}}{\norm{Z_M}_{p}},
	\label{eq:ind_norm}
\end{equation}
where the subscript $i$ stands for induced. In short, \eqref{eq:ind_norm} is a matrix form of representing the system gain considering a set of possible input signals $Z_M$. From the induced-2 norm, the following is obtained
\begin{equation}
\label{eq:gmi2}
    \norm{S_M}_{i2} = \bar{\sigma}(S_M) = \sqrt{\lambda_{max}\left(S_M' S_M\right)},
\end{equation}
where $\bar{\sigma}$ and $\lambda_{max}$ stands for largest singular value and largest eigenvalue, respectively, and comparing \eqref{eq:sinf_def} with \eqref{eq:ind_norm}, it can be seen that
\begin{equation}
    \label{eq:norm_Sinf}
    \norm{S}_\infty \approx \max_{Z_M \neq 0}\frac{\norm{S_M Z_M}_{2}}{\norm{Z_M}_{2}} = \sqrt{\lambda_{\max}(S_M' S_M)}.
\end{equation}

Since the $\norm{S(z,\rho)}_{\infty}$ norm can be estimated based on its IR via \eqref{eq:norm_Sinf}, an expression for the input signal $\zeta(k)$ and the output signal $\psi(k)$ of $S(z,\rho)$ has to be derived in order to estimate its impulse response in the VRFT design context.

\subsubsection{Input-output signals of the sensitivity transfer function}
\label{sssec:signals_S}
Considering the system presented in Figure~\ref{fig:vrft}, its sensitivity transfer function in \eqref{eq:sensitivity} can be rewritten as
\begin{equation}
\label{eq:S_dif}
    1 + C(z,\rho) G(z) = S^{-1}(z,\rho).
\end{equation}

Assuming that $u(k)$ is sufficiently informative to capture all relevant characteristics of $S(z,\rho)$, multiplying both sides of \eqref{eq:S_dif} by $u(k)$ achieves
\begin{equation}
\label{eq:S_halfway}
    u(k) + C(z,\rho) G(z) u(k) = S^{-1}(z,\rho) u(k).
\end{equation}
It is known that $G(z) u(k) = y(k)$. Substituting such relation in \eqref{eq:S_halfway}:
\begin{equation}
    u(k) + C(z,\rho) y(k) = S^{-1}(z,\rho) u(k).
\label{eq:S_full}
\end{equation}
From \eqref{eq:S_full}, the signals
\begin{equation}
\label{eq:S_signals}
    \xi(k) = u(k), \quad \zeta(k) = u(k) + C(z,\rho) y(k),
\end{equation}
can be derived. Expressions \eqref{eq:S_full} and \eqref{eq:S_signals} mean that when a signal $\zeta(k)$ formed by $u(k) + C(z,\rho) y(k)$ is applied to $S(z,\rho)$, an output $\xi(k) = u(k)$ is obtained. Therefore, the impulse response of $S(z,\rho)$ can be estimated considering the data set $\{ \xi,\zeta\}_{k=1}^{N}$.

In this work, the IR estimation is made through identification with regularization techniques since: i) the variance of the estimates increases with the order $M$, which is suppressed with the use of regularization \cite{Chen2012}; and ii) knowing that IR is a sparse signal for sufficiently high $M$, the use of regularization is known to improve sparse signal estimates \cite{Brunton2019}. The algorithm for regularized estimation of the impulse response is described in \cite{Chen2012,Chen2013}, and is available in MATLAB\textregistered~\cite{Matlab2017}, Python~\cite{Fiorio2021}, and R~\cite{Yerramilli2017}.

The inclusion of the $\Hinf$ constraint in the cost function spoils the convexity characteristic of the VRFT and the solution can no longer be obtained through the least squares algorithm. A strategy to deal with local minima and other characteristics that may arise from a non-convex cost function is to use metaheuristic optimization \cite{Du2016}. Nevertheless, the VRFT with robustness constraints is formally proposed in the next section, which also addresses the use of swarm intelligence algorithms to achieve a feasible solution.

\section{VRFT with robustness constraints}
\label{sec:method}
The proposed method regards a two-step procedure. The first step follows the design of a controller using the VRFT, as commented in Section~\ref{sec:vrft}. The same data from the first step is used in the second step. The need for a second experiment is avoided since the estimation of $M_S$, represented in this context as $\hat{M}_S (\rho)$, does not require new data, as derived in Subsection~\ref{ssec:ms_est}. In this case, the cost function $J^{VR}$ is modified by the addition of a robustness constraint, regarding the value of the estimated ($\hat{M}_S (\rho)$) and desired ($M_{Sd}$) $\norm{S(z,\rho)}_{\infty}$ norm, leading to a new optimization problem:
\begin{mini}|l|
  {\rho}{J^{VR}(\rho)}{}{}
  \addConstraint{\hat{M}_S(\rho) \leq M_{Sd}}{}{}
  ,
\end{mini}
which can be applied directly to the cost function as a penalty \cite{Luenberger2015}, resulting in the \textit{Swarm Intelligence} optimization cost function:
\begin{mini}|l|
  {\rho}{J^{SI} (\rho) = \norm{u(k) - C(z,\rho) (T_d^{-1}(z) - 1) y(k)}_{2}^{2} + c H(\rho)}{}{},
  \label{eq:opt_si}   
\end{mini}
where $c$ is a positive constant, usually with $c \gg 1$, and the $\Hinf$ penalty can be written as
\begin{equation}
\label{eq:P_leq}
    H(\rho) = \frac{1}{2} (max[0,\hat{M}_S(\rho) - M_{Sd}])^2. 
\end{equation}

The controller $C(z,\rho)$ in \eqref{eq:opt_si} can be chosen as belonging to any controller class $\mathcal{C}$, linearly parameterized in $\rho$ as $\mathcal{C} = \{ C(z,\rho), \rho \in \mathbb{R}^p \}$. The robustness index $\hat{M}_S(\rho)$ is estimated at each iteration of the swarm intelligence algorithm optimization following the procedure described in Subsection~\ref{ssec:ms_est}. 

Notice that an important insight can be drawn from \eqref{eq:opt_si} regarding the choice of $M_{Sd}$. If the choice is too ambitious, i.e., $M_{Sd}$ much lower than $\hat{M}_{S}(\rho_0)$ estimated with an initial VRFT-obtained controller with parameter $\rho_0$, the obtained parameter $\rho$ could be found too distant from the VRFT solution. Such a difference could considerably increase the value of $J^{VR}(\rho)$ to a point where the performance of the swarm intelligence-obtained controller is drastically affected. On the other hand, if the constant $c$ is chosen with an exorbitantly high value, only the cases where the optimization algorithm falls into a local minimum and is not able to minimize $H(\rho)$ would be affected, resulting in a final cost much higher than with a lower value of $c$. If the penalty element $H(\rho)$ is solved by the optimization algorithm, as seen in \eqref{eq:P_leq}, its value becomes zero, indifferent of the value of $c$, not affecting the VRFT cost $J^{VR}$.

Considering a search space $\mathcal{O} \in [l_b,u_b], l_b,u_b \in \mathbb{R}$, in order to accelerate the convergence of the metaheuristic algorithm, the initialization of the search agents can inherit the first step solution $\rho_0 \in \mathbb{R}^p$ as a central point, as expressed in
\begin{equation}
\label{eq:swarm_init}
    \overrightarrow{X}_{b}(0) = R \cdot \overrightarrow{X}(0) + \rho_0, \quad R = \frac{|\max \{ l_b, u_b \}|}{2},
\end{equation}
with $R$ being the initial population spawn radius, and $\overrightarrow{X}(0) \in \mathbb{R}^p$ a random position vector such that $\overrightarrow{X}(0) \sim U(0,1)$.

An inherent step of the method is to collect input-output data from the process, as suggested in \cite{Bazanella2014,Remes2021}. Remember to take into account system identification theory \cite{Ljung1999} for data to be sufficiently informative. Then, the two proposed design steps can be applied:
\begin{enumerate}
    \item Use the VRFT to design a controller for the process. Use a flexible reference model if the plant is NMP, as presented in \cite{Campestrini2011}. Check the obtained robustness index and proceed to the second step if it does not satisfy $\hat{M}_S \leq M_{Sd}$, or else, take the controller from Step 1 as final. Such a main step can be divided into the following specific steps:
    \begin{itemize}
        \item acquire a data set $\{ u,y\}_{k=1}^{N}$ from the closed-loop system with an initial stabilizing controller;
        \item use the data set to design a controller with the VRFT method, as detailed in Section~\ref{sec:vrft}. Controller parameters $\rho$ are obtained after the minimization procedure of the VRFT method. In the NMP case, parameters for the reference model $\hat{\eta}$ are also obtained;
        \item estimate the robustness index according to the method in Subsection~\ref{ssec:ms_est}. If $\hat{M}_S > M_{Sd}$ proceed to the second step, or else, use the VRFT-obtained controller with no further modification.
    \end{itemize}
    \item Apply a swarm intelligence algorithm, as presented next in Subsection~\ref{ssec:swarm}, considering the optimization problem described in \eqref{eq:opt_si} according to the desired value of $M_S$, with restriction applied in the form of a penalty as \eqref{eq:P_leq}, with the initial spawn of agents following the recommendation of \eqref{eq:swarm_init}. The second step can be divided in:
    \begin{itemize}
        \item implement the VRFT cost function with the penalty as in \eqref{eq:P_leq} regarding the desired maximum value of $M_S$;
        \item change the initialization procedure of the chosen swarm intelligence algorithm to consider a center spawn $\rho_0$, i.e., the VRFT-obtained solution at the first step, and a spawn radius as suggested in \eqref{eq:swarm_init} to accelerate convergence;
        \item execute the algorithm and obtain controller parameters that satisfy the robustness restriction.
    \end{itemize}
\end{enumerate}

The swarm intelligence algorithms considered for obtaining the results of this work are described next.

\subsection{Swarm intelligence algorithms}
\label{ssec:swarm}
Swarm intelligence algorithms consist of stochastic optimization algorithms that are based on the collective intelligence of groups composed of simple agents, usually based on the behavior of animals in nature \cite{Bonabeau2001}. In order to cope with the NFL theorems \cite{Wolpert:NFL:1997}, four algorithms are chosen to be used: Particle Swarm Optimization (PSO) \cite{Kennedy:PSO:1995} and Artificial Bee Colony (ABC) \cite{Karaboga2007}, which are well known and widely used \cite{Du2016,Talbi2009}; Grey Wolf Optimizer (GWO) \cite{Mirjalili:GWO:2014} and Improved Grey Wolf Optimizer (I-GWO) \cite{Nadimi:IGWO:2021}, more recent approaches with fewer hyperparameters than the aforementioned. In the following, the four algorithms are briefly presented.

\subsubsection{Particle Swarm Optimization} 
\label{ssec:pso}
Particle Swarm Optimization (PSO) involves populations (or swarms) in which each element is called a \textit{particle} that represents a form of directed mutation \cite{Talbi2009,Kennedy:PSO:1995}. The swarm is composed of $\ell$ particles searching in a $D$-dimensional space, which are initialized randomly within the search space with lower and upper bounds $l_b$ and $u_b$. Each particle has its own position ($\overrightarrow{X}_i$) and velocity ($\overrightarrow{V}_i$) and is considered as a possible solution for the problem. The best solution found locally by a particle $i$ is represented by $\overrightarrow{P}_i = \{ P_{i1},P_{i2},...,P_{iD} \}$, while $\overrightarrow{G} = \{ G_1,G_2,...,G_D \}$ is the best solution found globally. As for the standard algorithm, each particle is initialized at a random location with random velocity.

The pseudo-code for PSO is given in \ref{appendix:pso}, in which $max\_it$ is the maximum number of iterations and $f$ is the cost function. The constant $w_1$ is an inertia weight, $C_1$ is the cognitive learning factor, and $C_2$ is the social learning factor. Details for the implementation are available in the literature \cite{Kennedy:PSO:1995}.

\subsubsection{Artificial Bee Colony} 
\label{ssec:abc}
The Artificial Bee Colony (ABC) algorithm simulates the behavior of bees performed during their foraging process, conducting a local search in each iteration. Possible solutions are represented by food sources, while the quality of each solution is proportional to the nectar amount in each source \cite{Du2016,Karabog:ABC:2005,Karaboga2007}. There are three types of bees: scout, employed, and onlooker. At the initialization, the scout bees randomly find possible food sources (solutions). Each food source receives an employed bee. By roulette wheel selection, onlooker bees choose food sources to be exploited based on their quality, but both types perform local searches in their neighborhood.

The pseudo-code in \ref{appendix:abc} describes the ABC optimization procedure, where $\overrightarrow{X}_i$ with $D$ dimensions is the location of food sources. In this form, $\ell$ is the number of possible solutions (food sources) and the number of scouts, employed, and onlooker bees are taken as the same as the number of food sources. $L$ is the abandonment criteria, defined by the designer. Implementation details are available in \cite{Karabog:ABC:2005}.

\subsubsection{Grey Wolf Optimizer} 
\label{ssec:gwo}
The Grey Wolf Optimizer (GWO) is an algorithm based on the hunting behavior of grey wolves, which have a strict social dominant hierarchy. The leaders are the alphas ($\alpha$), who are responsible for making decisions. At the second level are the betas ($\beta$), subordinates to the alphas who help in decision-making and other pack activities. The third level wolves are the deltas ($\delta$), representing scouts, sentinels, elders, hunters, and caretakers. The rest of the pack is called omega, which must submit to the higher-ranking wolves \cite{Nadimi:IGWO:2021}. In the GWO algorithm, all wolves follow the mean position of the $\alpha$, $\beta$, and $\delta$s. At each iteration, the new three best wolves are re-defined, with the $\alpha$ position being the final solution. The pseudo-code for the GWO algorithm is presented in \ref{appendix:gwo}. Details for the implementation can be checked at \cite{Mirjalili:GWO:2014}.

\subsubsection{Improved Grey Wolf Optimizer}
\label{ssec:igwo}
There are three main problems noted in literature around the GWO algorithm \cite{Nadimi:IGWO:2021}: i) lack of population diversity; ii) imbalance between exploitation and exploration; iii) premature convergence. The Improved Grey Wolf Optimizer (I-GWO) tries to solve those issues by changing the search strategy of the GWO algorithm, including the Dimension Learning-based Hunting (DLH) strategy, which defines a new (possible) update position for each wolf based not just on the position of $\alpha$, $\beta$, and $\delta$ wolves, but also on the position of its neighbors. \ref{appendix:igwo} briefly describes the I-GWO. Details of the implementation are given at \cite{Nadimi:IGWO:2021}.

\section{Validation results}
\label{sec:examples}

In order to validate and illustrate the proposed method, two real-world-inspired examples are considered. The method is applied as suggested in Section~\ref{sec:method} with all four swarm intelligence algorithms commented in Subsection~\ref{ssec:swarm}. The results are compared in terms of: i) cost value obtained for best solution - best (lowest) cost; ii) $\norm{S(z,\hat{\rho})}_{\infty}$ value obtained for best solution; iii) convergence speed. Notice that the system model is only used to generate data in simulation. The knowledge of the model is neglected at any stage of the design, maintaining a pure data-driven fashion.

\subsection{Example 1: a second-order non-minimum phase plant}
\label{ssec:g1}

The first system to be considered is
\begin{equation}
\label{eq:g1}
    G_1(z) = \frac{-0.05(z-1.4)}{z^2 - 1.7z + 0.7325}
\end{equation}
with a time step of 1 second, which is similar to the discrete-time model of a Boost/Buck-Boost converter operating in continuous conduction mode, regarding the transfer function of the output voltage by the duty cycle \cite{Erickson2001}. The presence of a non-minimum phase zero makes it necessary to use the VRFT with a flexible criterion \cite{Bazanella2014} at the first step of the proposed method. 

\begin{figure} [!t]
	\centering
	\includegraphics[width=0.7\linewidth]{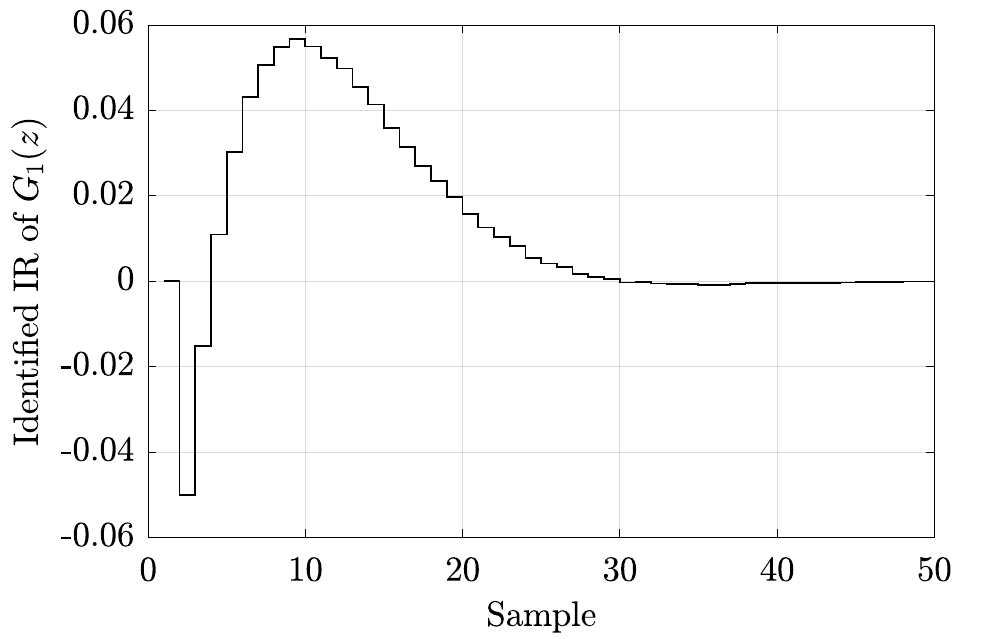} \\
	\caption{Identified impulse response of $G_1(z)$.}
	\label{fig:g1_ir}
\end{figure}

Assuming that the system model \eqref{eq:g1} is unknown, there is no previous knowledge about its zero being Non-Minimum Phase (NMP). In this sense, it is possible to analyze the estimated impulse response since the IR of NMP systems initially moves in the opposite direction (downwards) related to the steady-state one \cite{Brunton2019}. Therefore, a Pseudo-Random Binary Signal (PRBS), which is persistently exciting of high order \cite{Ljung1999}, containing $N=2000$ samples is applied to $G_1(z)$ in simulation, generating an output signal. Additive white Gaussian noise with a Signal-to-Noise Ratio (SNR) of 20~dB was added to the system at the output, representing measurement noise. With the input-output dataset, the IR of $G_1(z)$ can be identified with an IR identification algorithm available in the literature \cite{Chen2013,Fiorio2021,Matlab2017,Yerramilli2017}, resulting in the signal presented in Figure~\ref{fig:g1_ir}. Clearly, the IR initially goes downwards, indicating the presence of an NMP zero, justifying the VRFT with a flexible reference criterion \cite{Bazanella2014}.

\subsubsection{Data collection}
\label{sssec:g1_datacollection}
The data for estimation is obtained in a closed-loop with a proportional stabilizing controller \cite{Remes2021} since its presence in the system avoids signal divergence. By the small gain theorem \cite{Skogestad2005}, a stabilizing controller can be obtained following
\begin{equation}
    k_p < \frac{1}{\norm{G}_{\infty}}.
\end{equation}
Therefore, the stabilizing controller $k_p$ is chosen as
\begin{equation}
\label{eq:kp}
    k_p = \frac{0.5}{\norm{G_1(z)}_{\infty}} = 0.8039.
\end{equation}
In order to obtain the $\Hinf$ norm of $G_1(z)$, its impulse response is estimated according to \cite{Fiorio2021} and the norm is calculated as proposed in Subsection~\ref{ssec:ms_est}.

The excitation signal considered for the data acquisition is a PRBS with $N=2000$ samples. The signal is applied to the control reference of the closed-loop system formed by $G_1(z)$ with stabilizing controller $k_p$. The control output signal $u(k)$ and the system output signal $y(k)$ are acquired, forming the input-output set $\{ u,y\}_{k=1}^{N}$.

\subsubsection{Step 1 - VRFT with flexible criterion}
\label{sssec:vrft_g1}
Assume a situation where the control requirements are: i) null error in steady-state; ii) settling time approximately 2.5 times faster than in closed-loop with the stabilizing controller $k_p$; iii) null overshoot for a step reference. A reference model that meets such requirements, following the guidelines of \cite{Remes2021}, is chosen as
\begin{equation}
    T_d(z,\hat{\eta_0}) = \frac{-21 (z-1.01)}{(z-0.7)(z-0.3)}.
\end{equation}
Notice that the initial zero of $T_d(z)$ is set as greater than 1, as suggested in \cite{Bazanella2014}, allowing for the VRFT with a flexible criterion to identify the plant's NMP zero. The chosen controller class chosen to be used is the PID class of controllers, which gives
\begin{equation}
    \bar{C}(z) = 
    \left[
        1 \quad \frac{z}{z-1} \quad \frac{z-1}{z}
    \right]'
    .
\end{equation}

After solving the cost function \eqref{eq:vrft_j_f} according to the VRFT method with flexible criterion, the following solution pair for $\eta,\rho$ is obtained:
\begin{subequations}
    \begin{equation}
        \hat{\eta} = 
        \begin{bmatrix}
            -0.4793 & 0.6377
        \end{bmatrix}',
    \end{equation}
    \begin{equation}
        \hat{\rho} =        
        \begin{bmatrix}
            k_p & k_i & k_d
        \end{bmatrix}'
        =
        \begin{bmatrix}
            1.1246 & 0.3124 & 6.9713
        \end{bmatrix}',
        \label{eq:rho0}
    \end{equation}
\end{subequations}
resulting in a new reference model $T_d(z,\hat{\eta})$, and in the controller $C(z,\hat{\rho})$, respectively:
\begin{subequations}
\label{eq:vrft_sol_g1}
    \begin{equation}
        T_d(z,\hat{\eta}) = \hat{\eta} F(z) = \frac{-0.6899 (z - 1.33)}{(z-0.7)(z-0.2401)};
        \label{eq:vrft_td_g1}
    \end{equation}
    \begin{equation}
        C(z,\hat{\rho}) = \hat{\rho}' \bar{C}(z) = \frac{8.4083 (z^2 - 1.792z + 0.8291)}{z (z-1)}.
    \end{equation}
\end{subequations}
The non-dominant pole of the reference model, now $T(z,\hat{\eta})$, is updated together with the minimization of $\eta$ and $\rho$, as suggested in \cite{Remes2021}.


By estimating the $\Hinf$ norm of $S(z,\hat{\rho})$ of the closed-loop system based on the solution \eqref{eq:vrft_sol_g1}, an $\hat{M}_S = 2.1952$ is obtained, which may be too high for applications that require lower robustness indexes since it is greater than $2$ \cite{Skogestad2005}. The next subsection presents the application of the second step of the proposed method to reduce $M_S$ for the obtained VRFT solution.

\subsubsection{Step 2 - Swarm intelligence algorithm}
\label{sssec:swarm_g1}

The use of swarm intelligence algorithms for solving the proposed problem is straightforward. At their implementation - see \ref{appendix:pso}, \ref{appendix:abc}, \ref{appendix:gwo}, and \ref{appendix:igwo} for the pseudo-codes - the cost function $f$ is the proposed $J^{SI}(\rho)$ \eqref{eq:opt_si}. At each iteration, the input to $f$ is the parameter $\rho$, obtained in the previous iteration for each search agent since it is required to parameterize the controller $C(z,\rho)$ and to estimate the $\Hinf$ norm $\hat{M}_S(\rho)$.

The computation of $J^{SI}(\rho)$ is done by directly calculating the cost \eqref{eq:opt_si}. The norm $\hat{M}_S(\rho)$ can be obtained through \eqref{eq:norm_Sinf}, estimating the impulse response of $S(z,\rho)$ with the signals proposed in \eqref{eq:S_signals} and an IR estimation algorithm available in the literature, such as \cite{Fiorio2021}.

The upper search bound for the swarm intelligence algorithms is defined as $u_b = 10$, which should be sufficient considering that the maximum $\rho$ value of the VRFT-obtained controller is $6.9713$ and, taking into account a choice of $M_{Sd}$ that is not too ambitious, the resultant $\rho$ should not be too distant from the initialized value. The lower search bound is chosen as $l_b = 0$ to avoid negative controller gain, making the obtained controller passive \cite{Bao2007}.

The initialization of all agents is done randomly with a uniform distribution within the suggested spawn radius in \eqref{eq:swarm_init}, $R = u_b/2 = 5$, with the central point equal to the VRFT solution at Step 1 \eqref{eq:rho0}. The reference model for the cost function \eqref{eq:opt_si} is considered to be the VRFT with a flexible criterion reference model, $T_d(z,\hat{\eta)}$, obtained in Step 1 as \eqref{eq:vrft_td_g1}. The desired $M_S$ to be achieved, $M_{Sd}$, is set to 1.8, which is a sufficient value in terms of robustness, satisfies $M_{Sd} \leq 2$ and should not compromise substantially the performance of the system, which could happen if $M_{Sd} << 2$.

To make the comparison between algorithms possible, the number of agents was fixed to 50 and the number of iterations was limited to 100. In order to obtain a satisfactory number of realizations for the analysis of results each algorithm was executed 50 times. The hyperparameters for the PSO and ABC algorithms, except for the number of agents and the maximum number of iterations, are presented in Table~\ref{tab:parameters}. The PSO parameters were chosen as the MATLAB\textregistered\ default parameters of the Global Optimization Toolbox \cite{Matlab2017}, while the ABC parameters were used according to the algorithm implementation of \cite{Heris2015}. GWO and I-GWO do not contain any hyperparameter aside from the number of agents and the maximum number of iterations. 

\begin{table}[!t]
	\centering
	\begin{tabular}{c c c}
		\hline
	    Algorithm & Parameter settings & Value \\
		\hline
		\multirow{3}{*}{PSO} & Cognitive learning factor ($C_1$) & 1.49 \\
		                     & Social learning factor ($C_2$) & 1.49 \\
		                     & Inertia range (range of $w_1$) & [0.1,1.1] \\
		                     
		\cline{2-3}
        \multirow{2}{*}{ABC} & Abandonment criteria ($L$) & 90 \\
        		             & Acceleration coefficient ($a$) & 1 \\
		\hline
	\end{tabular}
	\caption{Parameters settings for PSO and ABC.}		
	\label{tab:parameters} 
\end{table}

\begin{figure} [!t]
	\centering
	\includegraphics[width=0.7\linewidth]{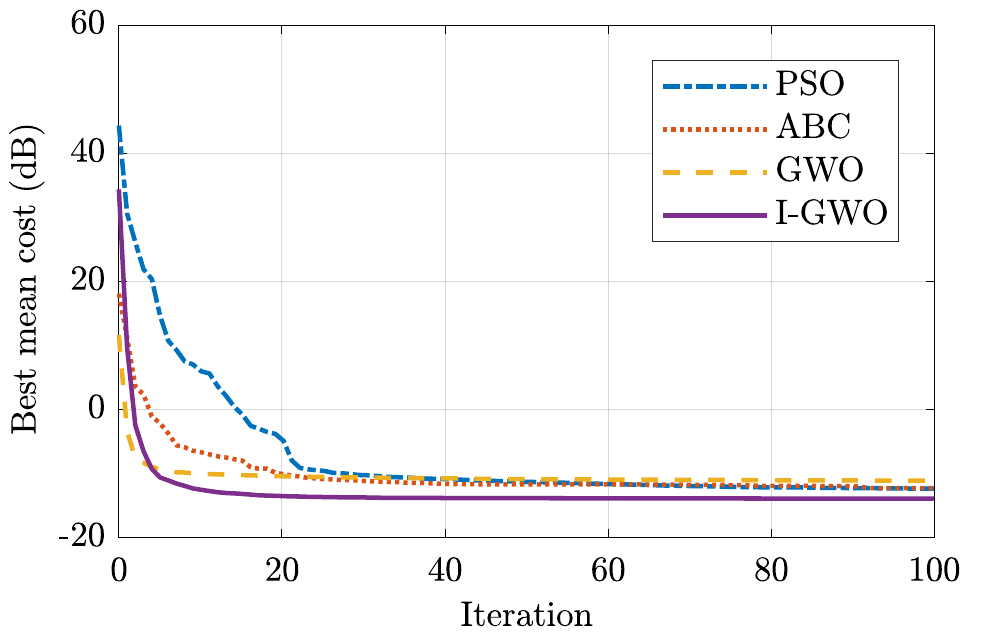} \\
    \caption{Average convergence curves for all algorithms considering a Monte Carlo experiment of 50 executions for example 1.}
	\label{fig:conv_g1}
\end{figure}

Figure~\ref{fig:conv_g1} shows the average convergence curve of all algorithms for 50 runs, considering the system $G_1(z)$ as aforementioned. Table~\ref{tab:timing_g1} presents the time that each iteration took and the number of iterations to converge, considering the average value for all 50 realizations and a convergence criterion of $\delta = 1 \times 10^{-3}$ from an iteration cost to the subsequent one. The results were obtained with an Intel Core i5 4670 3.40 GHz processor. The I-GWO algorithm took a longer time to converge, followed by ABC, PSO, and finally, GWO. Comparing similar algorithms, I-GWO obtained a duration for one iteration that is more than twice the same duration for the GWO. But since the optimization of the proposed cost function is executed \textit{offline}, where the obtained parameters ($\rho$) are applied to the controller without any further modification, this is not considered an issue for the current application.

\begin{figure} [!t]
	\centering
	\includegraphics[width=0.7\linewidth]{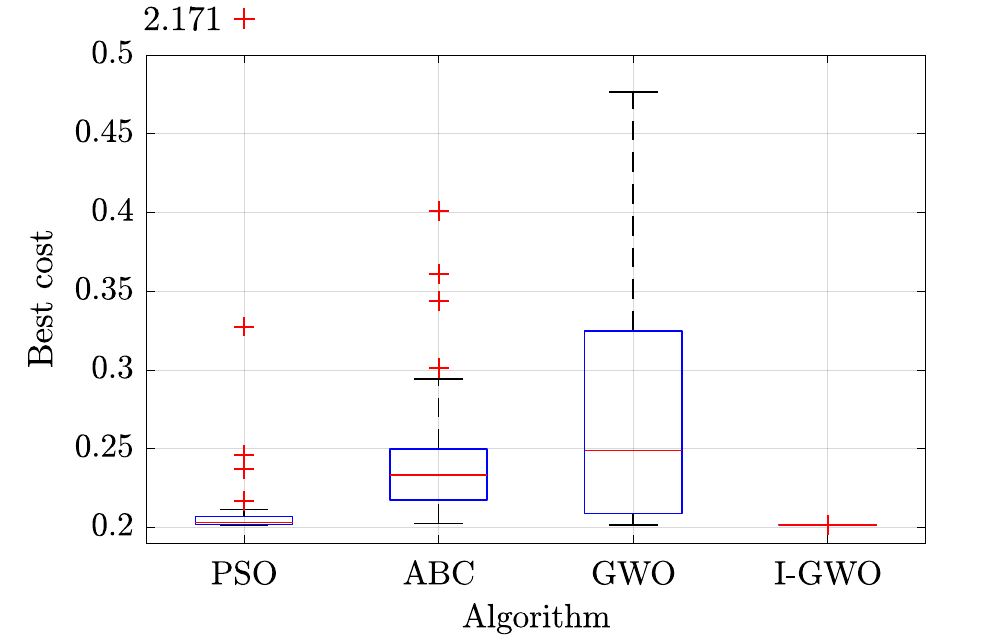} \\
	\caption{Box plot of a Monte Carlo experiment with 50 executions for all algorithms in terms of best cost value obtained for example 1.}	
	\label{fig:fitness_box_g1}
\end{figure}

\begin{figure}[!t]
    \centering
      \begin{minipage}{0.49\linewidth}
        \centering
        \includegraphics[width=1.0\linewidth]{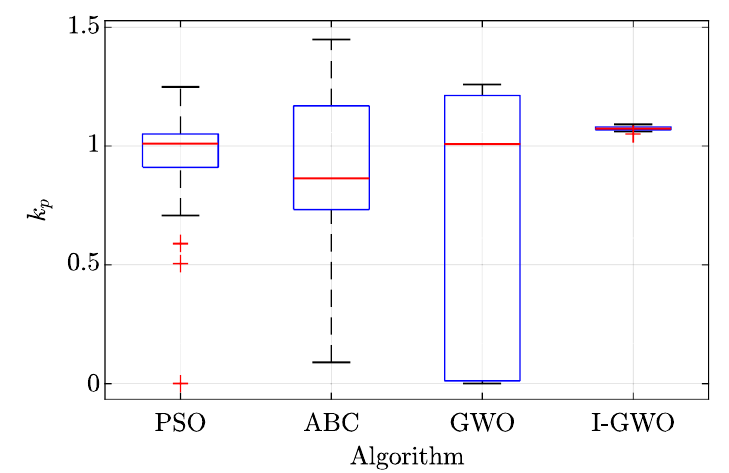}
        \subcaption{$k_p$} 
      \end{minipage} 
      \begin{minipage}{0.49\linewidth}
        \centering
        \includegraphics[width=1.0\linewidth]{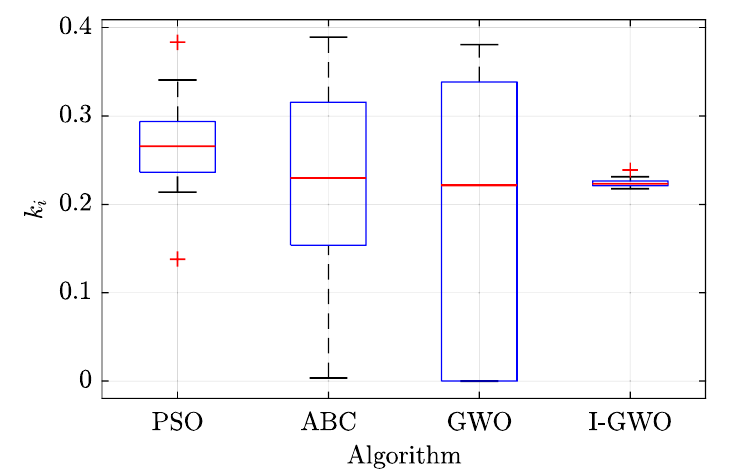}
        \subcaption{$k_i$} 
      \end{minipage} 
      \centering
      \begin{minipage}{0.49\linewidth}
        \centering
        \includegraphics[width=1.0\linewidth]{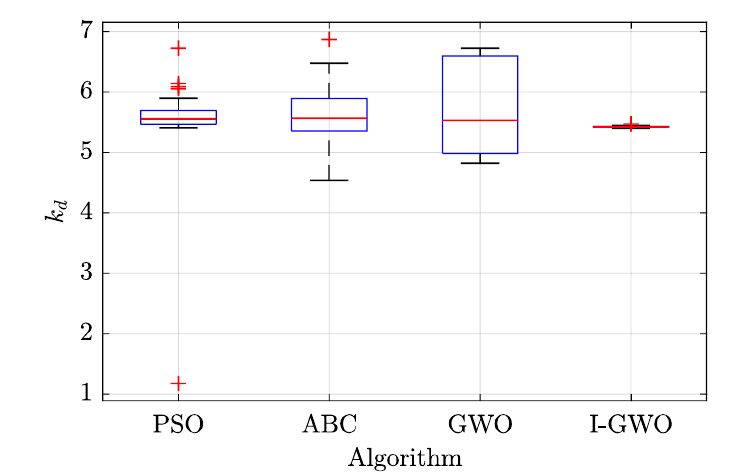}
        \subcaption{$k_d$} 
      \end{minipage}
    \caption{Box plot of the obtained $k_p$ ($\rho_1$), $k_i$ ($\rho_2$), and $k_d$ ($\rho_3$) controller parameters for all algorithms evaluated at 50 executions for example 1.}
\label{fig:g1k}
\end{figure}

Considering all 50 executions per swarm intelligence algorithm, Figure~\ref{fig:fitness_box_g1} presents the best (lower) cost statistics for all algorithms at example 1 in the form of a box plot. Clearly, I-GWO achieved the most desirable performance in terms of cost since it contains fewer outliers and a very low dispersion if compared to the other algorithms' solutions. PSO, ABC, and GWO, in general, result in higher cost values than I-GWO for the considered cost function. Table~\ref{tab:fitness_quant_g1} shows the quantitative values related to the best cost of all algorithms at each run, confirming the conclusions taken from Figure~\ref{fig:fitness_box_g1}. The obtained values for the parameter $\rho = [k_p \ k_i \ k_d]'$ are presented in Figure~\ref{fig:g1k} and show how the lower cost dispersion obtained with the I-GWO algorithm can influence the values of the controller parameters and their tendency. For the case of example 1, the I-GWO is the most robust algorithm in relation to the random initialization of the parameters. Such a characteristic is desired in a real-world scenario, reducing the number of trials in the design procedure to obtain a satisfactory solution. Also, notice that the median solution obtained with the I-GWO is not too far from the VRFT solution \eqref{eq:rho0} obtained at Step 1, which agrees with the previous hypothesis that a value of desired robustness $M_{Sd}$ chosen as not too ambitious when compared with the initial VRFT solution \eqref{eq:rho0} should result in a final solution that is close to the initial (VRFT) one.

\begin{table}[!t]
	\centering
	\begin{tabular}{c c c c c}
		\hline
		Algorithm & 1-it. time (s) & It. to converge & Time to converge (s) \\
		\hline
		PSO & $\ 8.04$ & $41$ & $329.75$ \\
		ABC & $22.63$ & $19$ & $430.02$ \\
		GWO & $11.25$ & $14$ & $157.53$ \\
		I-GWO & $24.18$ & $20$ & $483.66$ \\
		\hline		
	\end{tabular}\\
	\caption{Time for convergence of all algorithms for example 1.}	
	\label{tab:timing_g1} 
\end{table}
\begin{table}[!t]
	\centering
	\begin{tabular}{c c c c c}
		\hline
		Algorithm & median & $\sigma$ & min & max \\
		\hline
		PSO & $0.2032$ & $0.2782$ & $0.2017$ & $2.1710$ \\
		ABC & $0.2334$ & $0.0402$ & $0.2025$ & $0.4009$ \\
		GWO & $0.2489$ & $0.0858$ & $0.2017$ & $0.4764$ \\
		I-GWO & $0.2017$ & $5.0516 \times 10^{-5}$ & $0.2017$ & $0.2018$ \\
		\hline		
	\end{tabular}
	\caption{Quantitative results from the box plot in terms of best cost for example 1.}		
	\label{tab:fitness_quant_g1} 
\end{table}

\begin{figure}[!t]
	\centering
	\includegraphics[width=0.7\linewidth]{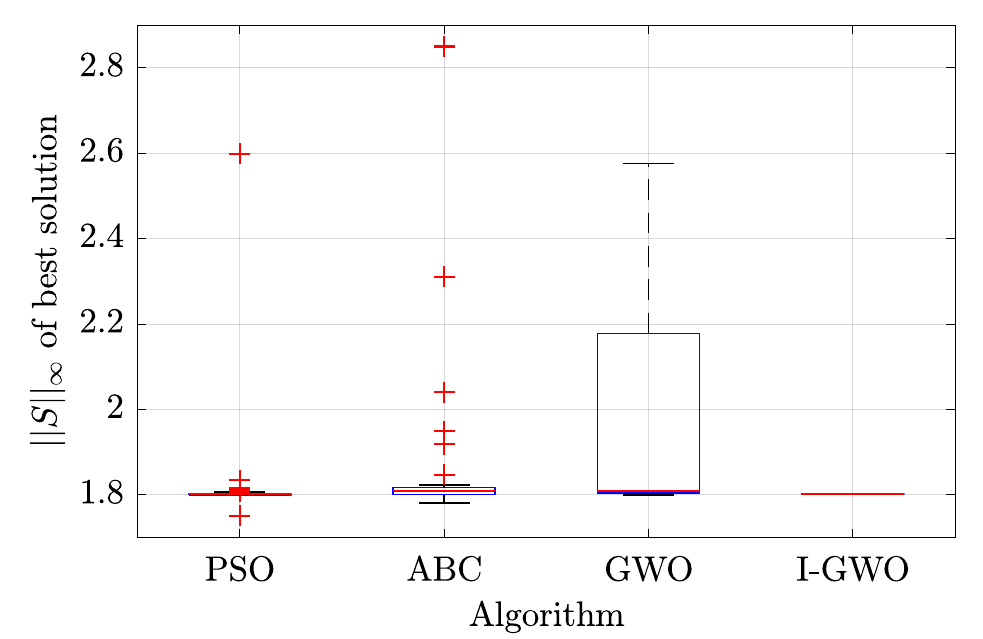} \\
	\caption{Box plot of 50 executions for all algorithms in terms of $\norm{S(z,\hat{\rho})}_{\infty}$ value obtained for example 1.}	
	\label{fig:sinf_box_g1}
\end{figure}

Figure~\ref{fig:sinf_box_g1} presents the box plot for the obtained $\norm{S(z,\hat{\rho})}_{\infty}$ by the best solution of each algorithm at each run, in a closed-loop with $G_1(z)$. I-GWO obtained the most desired result in terms of $\hat{M}_S$ considering the lack of outliers and the low dispersion. PSO had one outlier with $\hat{M}_S > 2$, while ABC obtained three outliers of higher $\hat{M}_S$, and the performance by the GWO algorithm for this problem was not satisfactory since there were too many solutions that achieved an $\hat{M}_S$ higher than 2. Table~\ref{tab:sinf_quant_g1} shows the quantitative data of the box plot presented in Figure~\ref{fig:sinf_box_g1}, in agreement with what is commented over the results.
\begin{table}[!t]
	\centering
	\begin{tabular}{c c c c c}
		\hline
		Algorithm & median & $\sigma$ & min & max \\
		\hline
		PSO & $1.8014$ & $0.1130$ & $1.7509$ & $2.5988$ \\
		ABC & $1.8091$ & $0.4740$ & $9.9798 \times 10^{-5}$ & $2.8496$ \\
		GWO & $1.8087$ & $0.2923$ & $ 1.7995$ & $2.5760$ \\
		I-GWO & $1.8020$ & $4.5561 \times 10^{-4}$ & $1.8008$ & $1.8030$ \\
		\hline		
	\end{tabular}
	\caption{Quantitative results from the box plot in terms of $\norm{S(z,\hat{\rho})}_{\infty}$ for example 1.}
	\label{tab:sinf_quant_g1} 
\end{table}

\begin{figure}[!htb]
	\centering
	\includegraphics[width=0.7\linewidth]{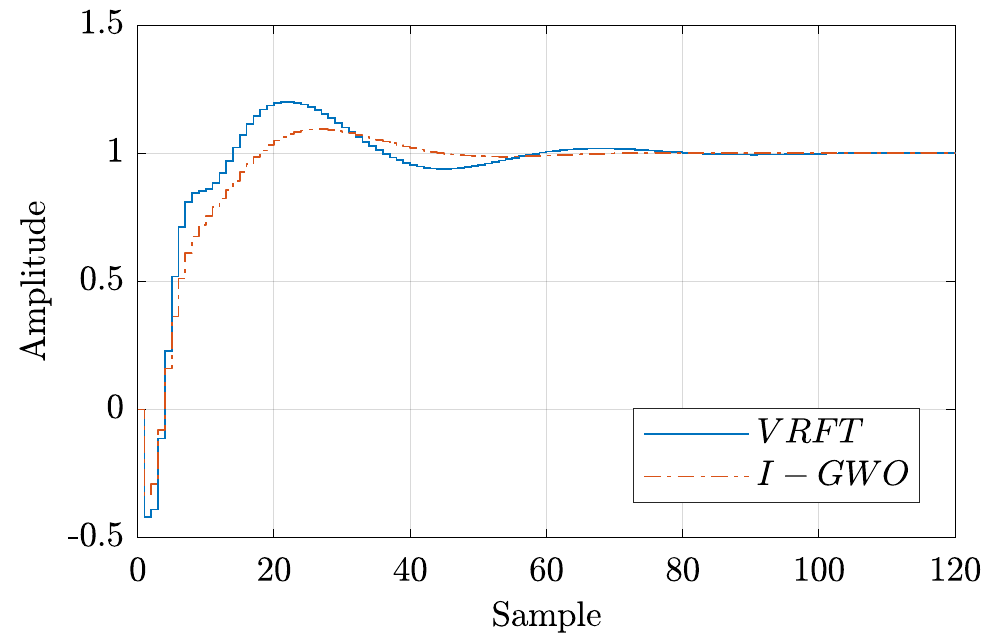} \\
	\caption{Output signal for a reference step signal applied to the controlled plant with a controller designed with the proposed method using I-GWO and the VRFT-designed controller for example 1.}	
	\label{fig:g1_step}	
\end{figure}

To briefly demonstrate that, although the robustness is increased with the proposed method, the reference tracking is not overly penalized, a step reference was applied to the controlled plant of example 1 using a controller designed by the proposed method with the I-GWO algorithm, with parameters $\rho_{IGWO} = [1.090 \quad 0.2194 \quad 5.4018]$, and the VRFT controller \eqref{eq:rho0} obtained at step 1. As shown in Figure~\ref{fig:g1_step}, the VRFT-controlled system achieves a settling time of 54 seconds, a step overshoot of 20 \% and an undershoot of 42 \%. For the proposed method, a settling time of 39 seconds is achieved, with 9 \% of overshoot and 33 \% of undershoot. Notice that, in this specific case, the proposed method could even enhance the reference tracking performance in terms of settling time, however, this is not expected in most of the systems since the inclusion of a robustness constraint will penalize the VRFT cost function. The reduced overshoot and undershoot are a consequence of the increased robustness, with an $\hat{M}_S$ of 2.2079 for the VRFT and 1.8030 for the I-GWO-designed controller.

\subsection{Example 2: fourth-order plant}
\label{ssec:g2}
The fourth-order plant consists of 
\begin{equation}
    G_2(z) = \frac{0.1381 (z-0.95) (z^2 - 1.62 z + 0.6586)}                                   {(z^2 - 1.7z + 0.7325) (z^2 - 1.84z + 0.8564)},
\end{equation}
with a time step of 1 second, which has the same structure as the discrete-time transfer function of a SEPIC converter, from duty cycle to output voltage \cite{Kassick2011}. Since the plant's zeros have minimum phase, which can be evaluated with data as aforementioned in Subsection~\ref{ssec:g1}, the VRFT method is used without the flexible reference model \cite{Bazanella2014}.

\subsubsection{Data collection}
\label{sssec:g2_datacollection}

For plant $G_2(z)$, the data is obtained in the same way as described, for example 1, in Subsection~\ref{ssec:g1}, with a PRBS signal of $N=2000$ samples applied to the closed-loop system with stabilizing controller
\begin{equation}
    k_p = \frac{0.5}{\norm{G_2(z)}_{\infty}} = 0.3828,
\end{equation}
considering additive white Gaussian noise with an SNR of 20~dB to represent measurement noise. The input-output set is formed by $\{ u,y\}_{k=1}^{N}$.

\subsubsection{Step 1 - VRFT}

After the data is acquired, the next step is to use the VRFT to design a controller, which solves the cost function \eqref{eq:vrft_j}. For this example, the following control requirements are assumed: i) null steady-state error; ii) settling time of approximately 6.5 times faster than the closed-loop settling time with stabilizing controller $k_p$; iii) null overshoot for a step reference. Considering such requirements, the choice of the reference model is made as suggested in \cite{Remes2021}, obtaining
\begin{equation}
    Td(z) = \frac{1.4 (z-0.6)}{(z-0.3) (z-0.2)}.
\end{equation}

Supposing a limited situation where only a PI controller is available, e.g., because of hardware limitations on a certain product. Therefore, the controller class to be considered is the PI class of controllers, resulting in
\begin{equation}
    \bar{C}(z) = \left[ 1 \quad \frac{z}{z-1} \right]'.
\end{equation}
The obtained VRFT solution results in the controller parameter
\begin{equation}
\label{eq:vrft_sol_ex2}
    \hat{\rho} = [k_p \quad k_i] = [6.6568 \quad 3.3728].
\end{equation}
Via \eqref{eq:c-rhobar}, the controller
\begin{equation}
    C(z,\hat{\rho}) = \hat{\rho}' \bar{C}(z) = \frac{10.03 (z-0.6637)}{(z-1)}
\end{equation}
is obtained.

Considering the VRFT-obtained solution \eqref{eq:vrft_sol_ex2}, the robustness index of the system can be estimated according to Subsection~\ref{sec:Ms}, obtaining $\hat{M}_S = 2.2767$. As aforementioned, an $M_S \leq 2$ is desired to ensure sufficient robustness \cite{Skogestad2005}, which leads to the application of the second step of the proposed method.

\subsubsection{Step 2 - Swarm intelligence algorithm}

The swarm intelligence algorithms PSO, ABC, GWO, and I-GWO are applied to the problem \eqref{eq:opt_si} for the fourth-order plant case, according to the pseudo-codes presented in \ref{appendix:pso}, \ref{appendix:abc}, \ref{appendix:gwo}, and \ref{appendix:igwo}, where $f$ is the cost function \eqref{eq:opt_si}, as already commented at Subsection~\ref{sssec:swarm_g1}. The upper search bound is kept at $u_b = 10$ and the lower bound at $l_b = 0$, in order to increase the passivity of the controller as mentioned in Subsection~\ref{ssec:g1}. An upper bound of $10$ should be sufficient, considering that the maximum desired robustness is not too far from the estimated robustness index at the end of step 1. The initial population spawn radius follows \eqref{eq:swarm_init}, $R = (|u_b|+|l_b|)/2 = 5$, with the solution of the VRFT at Step 1 \eqref{eq:vrft_sol_ex2} as the central point. The desired $\norm{S(z,\hat{\rho})}_{\infty}$ is set to 1.5, which satisfies $M_{Sd} \leq 2$ and is not much lower than 2.

The number of agents of all algorithms is set to 50, with a maximum of 100 iterations per execution. Each algorithm is executed 50 times for different noise realizations, such that the results are sufficiently representative for further analysis. For PSO and ABC algorithms, parameters are set as presented in Table~\ref{tab:parameters}. Aside from the number of agents and the maximum number of iterations, no other parameter is set by the user with the proposed GWO and I-GWO algorithm.

\begin{figure} [!t]
	\centering
	\includegraphics[width=0.7\linewidth]{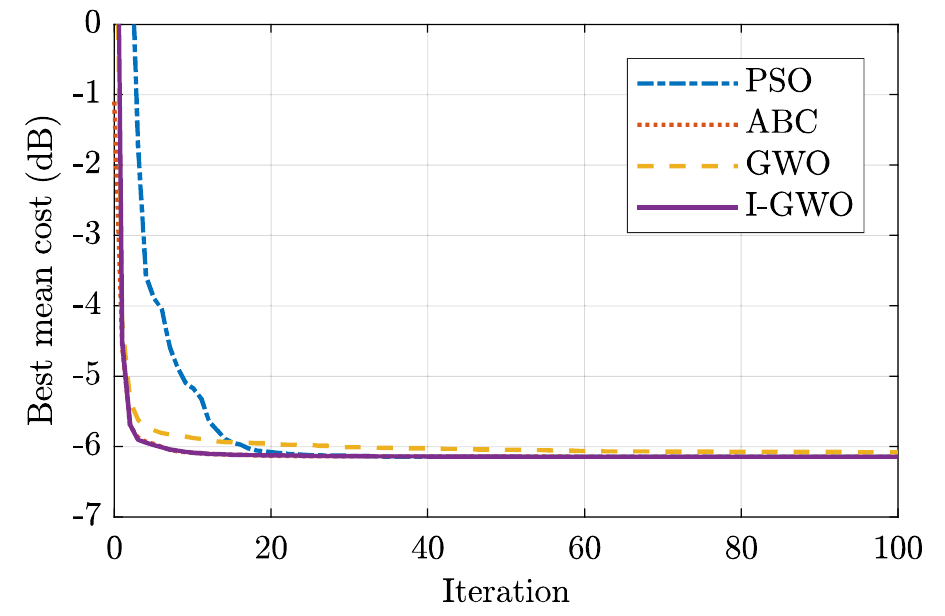} \\
	\caption{Average convergence curves for all algorithms considering a Monte Carlo experiment of 50 executions for example 2.}
	\label{fig:conv_g2}
\end{figure}

The average convergence curve of all algorithms for this case is presented in Figure~\ref{fig:conv_g2}. Table~\ref{tab:timing_g2} presents the time for one iteration and the number of iterations each algorithm took to converge, considering a convergence criterion $\delta = 1 \times 10^{-3}$. The hardware configuration is the same as in example 1. The ABC algorithm was the slowest algorithm in this example, followed by I-GWO, PSO, and at last, GWO. In comparison to example 1, GWO is still the fastest algorithm to converge, while ABC took I-GWO place as the slowest convergence. 

\begin{figure} [!t]
	\centering
	\includegraphics[width=0.7\linewidth]{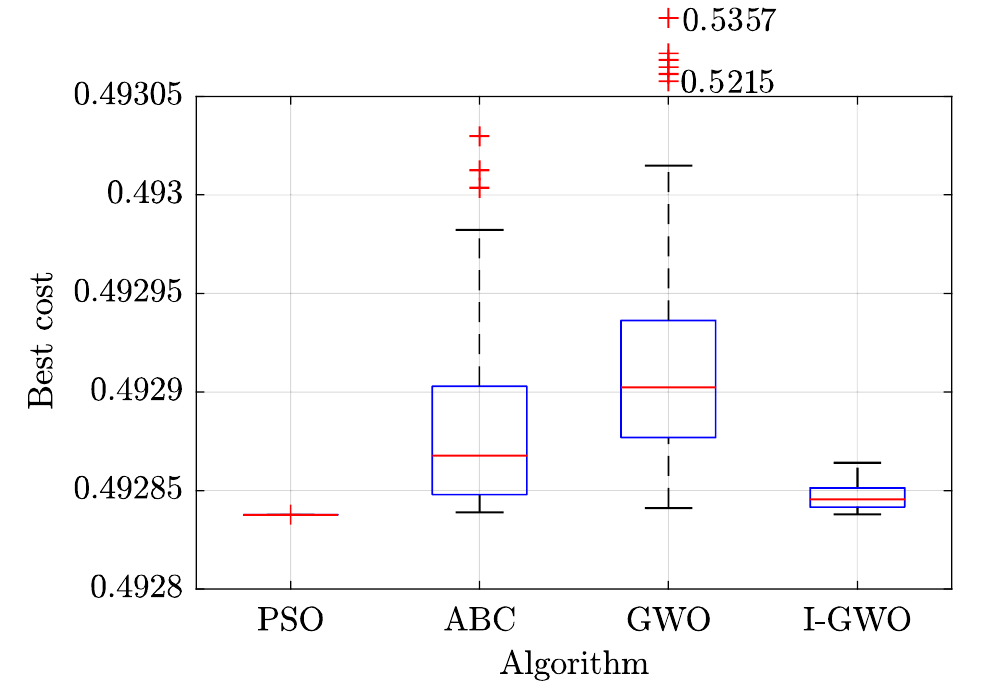} \\
	\caption{Box plot of a Monte Carlo experiment with 50 executions for all algorithms in terms of best cost value obtained for example 2.}	
	\label{fig:fitness_box_g2}
\end{figure}

Figure~\ref{fig:fitness_box_g2} shows the box plot regarding the best fitness value for each algorithm, considering all executions. PSO, ABC, and I-GWO did not present far outliers, as those seen in the case of GWO. The quantitative values of the box plot are shown in Table~\ref{tab:fitness_quant_g2}. In this case, the most desirable result is obtained by the PSO algorithm. Although, notice that the scale of the y-axis (best cost) only varies at the third decimal place, which means that PSO, ABC, and I-GWO solutions should perform very similarly in practice.
\begin{table}[!t]
	\centering
	\begin{tabular}{c c c c c}
		\hline
		Alg. & 1-it. time (s) & It. to converge & Time to converge (s) \\
		\hline
		PSO & $7.24$ & $20$ & $144.72$ \\
		ABC & $22.93$ & $10$ & $229.32$ \\
		GWO & $7.55$ & $9$ & $68.00$ \\
		I-GWO & $22.34$ & $9$ & $201.10$ \\
		\hline		
	\end{tabular}\\
	\caption{Time for convergence of all algorithms for example 2.}	
	\label{tab:timing_g2} 
\end{table}
\begin{table}[!t]
	\centering
	\begin{tabular}{c c c c c}
		\hline
		Algorithm & median & $\sigma$ & min & max \\
		\hline
		PSO & $0.49284$ & $2.2270 \times 10^{-9}$ & $0.49284$ & $0.49284$ \\
		ABC & $0.49287$ & $4.7174 \times 10^{-5}$ & $0.49284$ & $0.49303$ \\
		GWO & $0.49290$ & $1.0414 \times 10^{-2}$ & $0.49284$ & $0.53568$ \\
		I-GWO & $0.49285$ & $6.4636 \times 10^{-6}$ & $0.49284$ & $0.49286$ \\
		\hline	
	\end{tabular}
	\caption{Quantitative results from the box plot in terms of best cost for the example with system $G_2(z)$.}	
	\label{tab:fitness_quant_g2}
\end{table}

\begin{figure}[!t]
	\centering
	\includegraphics[width=0.7\linewidth]{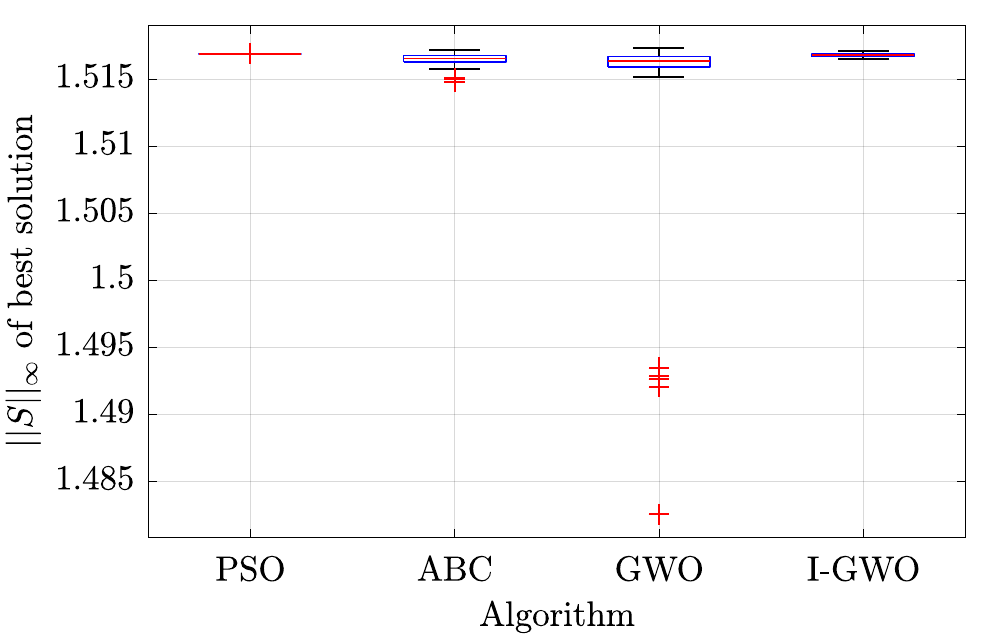} \\
	\caption{Box plot of a Monte Carlo experiment with 50 executions for all algorithms in terms of $\norm{S(z,\hat{\rho})}_{\infty}$ for example 2.}	
	\label{fig:sinf_box_g2}
\end{figure}

The $\norm{S(z,\hat{\rho})}_{\infty}$ norm obtained for the best solution at each run is shown in Figure~\ref{fig:sinf_box_g2}, with its quantitative values presented in Table~\ref{tab:sinf_quant_g2}. PSO, ABC, and I-GWO present very similar values for the obtained $\Hinf$ norm, which shows that the performance of those solutions is close, as also noticed from the cost analysis. GWO is the only algorithm that presents outliers in this case, which, in comparison to I-GWO, indicates that the local minima problems pointed out in literature \cite{Nadimi:IGWO:2021} indeed can occur.

\begin{figure}[!t]
    \centering
      \begin{minipage}{0.49\linewidth}
        \centering
        \includegraphics[width=1.0\linewidth]{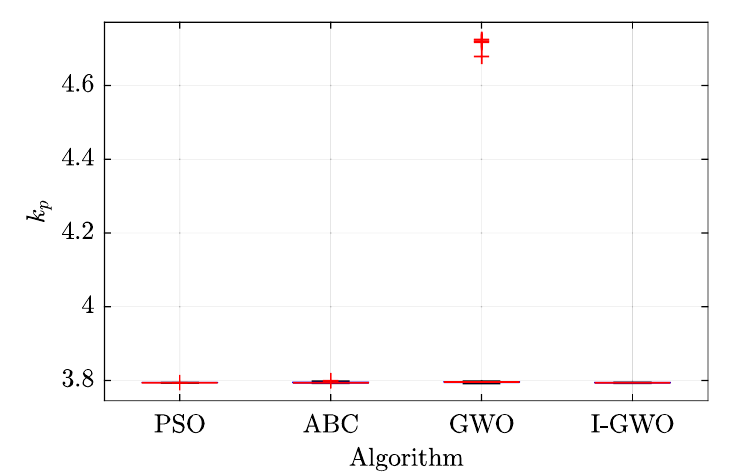} 
        \subcaption{$k_p$} 
      \end{minipage} 
      \begin{minipage}{0.49\linewidth}
        \centering
        \includegraphics[width=1.0\linewidth]{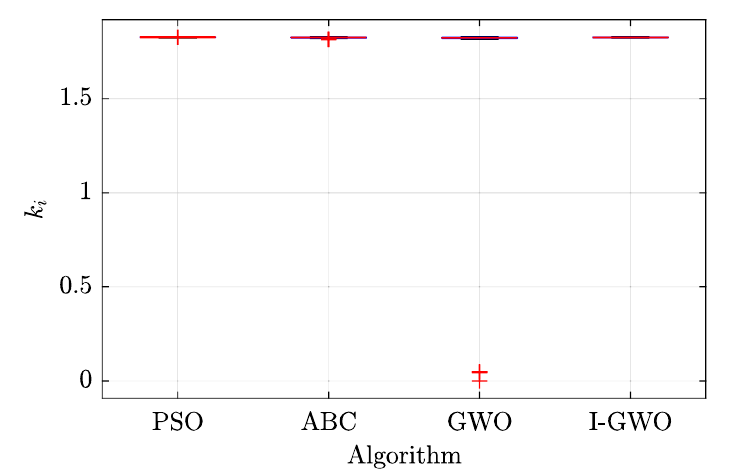} 
        \subcaption{$k_i$}
      \end{minipage} 
    \caption{Box plot of the obtained $k_p$ ($\rho_1$) and $k_i$ ($\rho_2$) controller parameters for all algorithms evaluated at 50 executions for example 2.}
\label{fig:g2k}
\end{figure}

Additionally, similar conclusions can be drawn from the controller parameter's values obtained with each optimization algorithm for the proposed problem. In Figure~\ref{fig:g2k}, outliers are only obtained with the GWO algorithm, where the gain of the integral part of the controller, $k_i$, achieves values close to zero, which has, as a consequence, a lower performance for the controlled system, even though the robustness criteria is met. The obtained $\rho$ for PSO, ABC, and I-GWO are very similar for all executions, indicating that any of the three algorithms could be used for this example without significant loss of performance or robustness when compared to one another.

\begin{table}[!t]
	\centering
	\begin{tabular}{c c c c c}
		\hline
		Algorithm & median & $\sigma$ & min & max \\
		\hline
		PSO & $1.5169$ & $1.4558 \times 10^{-6}$ & $1.5169$ & $1.5169$ \\
		ABC & $1.5165$ & $4.4060 \times 10^{-4}$ & $1.5148$ & $1.5172$ \\
		GWO & $1.5133$ & $8.4447 \times 10^{-3}$ & $1.4825$ & $1.5173$ \\
		I-GWO & $1.5168$ & $1.2216 \times 10^{-4}$ & $1.5165$ & $1.5171$ \\
		\hline	
	\end{tabular}
	\caption{Quantitative results from the box plot in terms of $\norm{S(z,\hat{\rho})}_{\infty}$ for example 2.}	
	\label{tab:sinf_quant_g2} 
\end{table}

\begin{figure}[!htb]
	\centering
	\includegraphics[width=0.7\linewidth]{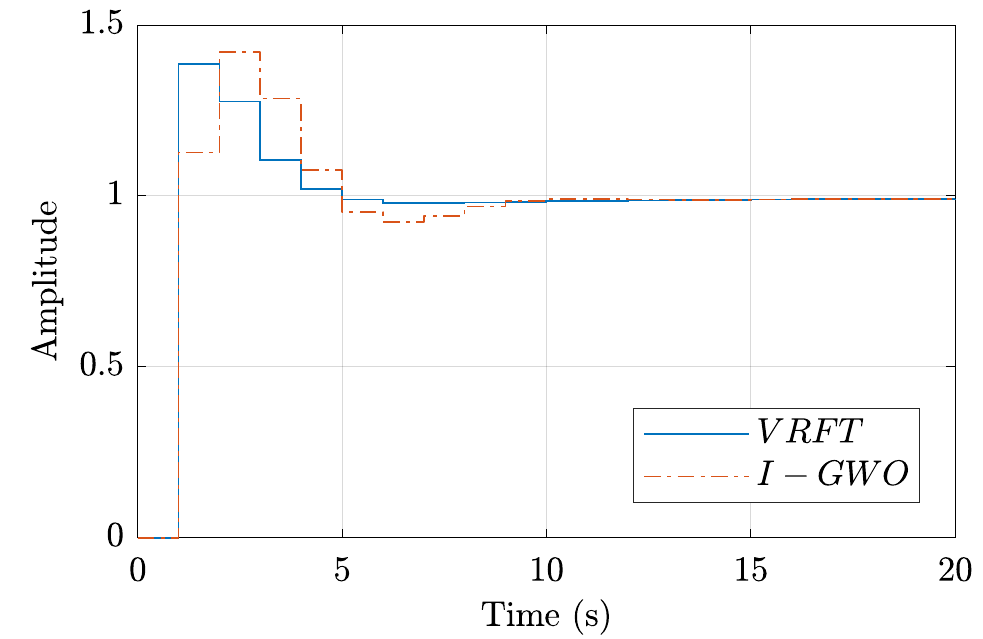} \\
	\caption{Output signal for a reference step signal applied to the controlled plant with a controller designed with the proposed method using I-GWO and the VRFT-designed controller for example 2.}	
	\label{fig:g2_step}	
\end{figure}

In the same way as for example 1, a step reference is applied to the controlled plant of example 2 using a VRFT-designed controller and a controller designed with the proposed method using I-GWO, as shown in Figure~\ref{fig:g2_step}. In this case, as one would expect, the reference tracking was penalized in terms of settling time, from 5 seconds (VRFT) to 14 seconds (I-GWO-designed). The overshoot is 42 \% for the VRFT controller and 29 \% for the proposed method with I-GWO, reflecting the reduced $\hat{M}_S$ - from 2.3834 to 1.5156.

\newpage
\section{Conclusion}
\label{sec:conclusion}

This work proposed a data-driven one-shot technique to increase the robustness of a closed-loop discrete-time system by changing the controller parameters using swarm intelligence algorithms. The considered optimization problem \eqref{eq:opt_si} is the VRFT cost function with the addition of a penalty regarding the value of the $\norm{S(z,\rho)}_{\infty}$ norm, which can be directly used as a measure of robustness. Such a value is estimated via an impulse response at each iteration of the metaheuristic algorithm.

Four swarm intelligence algorithms - PSO, ABC, GWO, and I-GWO - have been considered to illustrate the proposed technique with two real-world inspired plants. For the example of a second-order non-minimum phase plant, I-GWO obtained the best results in terms of dispersion, outliers, and cost, with acceptable values of $\norm{S(z,\rho)}_{\infty}$. PSO achieved similar results in this case, while ABC and GWO algorithms had a higher occurrence of outliers - in terms of cost and $\hat{M}_S$. In the case of the fourth-order minimum-phase plant, ABC, I-GWO, and PSO obtained very similar results, all satisfactory in terms of cost and $\Hinf$ norm of $S(z,\rho)$. GWO, on the other hand, obtained a few outliers that, although the criterion of robustness was met, those cases achieved a very low integral gain, which could drastically affect the performance of the system. In terms of speed of convergence, I-GWO and ABC were the slowest algorithms, while PSO converged faster and GWO was the fastest.

As for future works, it is suggested: the inclusion of other constraints (e.g., for control effort) simultaneously with the robustness constraints; the use of other types of metaheuristics, as evolutionary of physics-based algorithms; the inclusion of a robustness constraint to the OCI, VDFT, or DD-LQR methods; extension of the current work for MIMO systems.

\section{Acknowledgments}
\label{sec:acknowledgments}
This study was financed in part by the Coordenação de Aperfeiçoamento de Pessoal de Nível Superior - Brasil (CAPES) - Finance Code 001, partly by the Fundação de Amparo à Pesquisa e Inovação do Estado de Santa Catarina (FAPESC) - Grant number 288/2021, and partly by Conselho Nacional de Desenvolvimento Científico e Tecnológico - CNPq - Brazil.

\appendix
\section{}
\label{appendix:pso}
\begin{algorithm}[H]
\SetKwInOut{Input}{input}\SetKwInOut{Output}{output}
  \Input{$f$, $l_b$, $u_b$, $\ell$, $max\_it$, $w_1$, $C_1$, $C_2$}
  \Output{$\overrightarrow{G}$}
  initialize $\ell$ particles with random position $\overrightarrow{X}_i(0)$ within $l_b$, $u_b$, and velocity $\overrightarrow{V}_i(0)$\;
  define (initial) best local solution $\overrightarrow{P}_i$ as $f[\overrightarrow{X}_i(0)]$ for each particle\;
  define (initial) best global solution $\overrightarrow{G}$ as the position of the particle with the best fitness\;
  \For{$n \in \{1,...,max\_it\}$}{
    \For{$i \in \{1,...,\ell \}$}{
        define new velocities $\overrightarrow{V}_i(n)$ for each particle with $w_1$, $C_1$, $C_2$\;
        define new positions $\overrightarrow{X}_i(n)$ for each particle\;
      }
    update best local solution for all particles\;      
    $\overrightarrow{G} \gets$ position of the particle with higher fitness
  }
  \caption{Particle swarm optimization pseudo-code}
\label{alg:pso}  
\end{algorithm}

\section{}
\label{appendix:abc}
\begin{algorithm}[H]
\SetKwInOut{Input}{input}\SetKwInOut{Output}{output}
    \Input{$f$, $l_b$, $u_b$, $\ell$, $max\_it$, $L$}
    \Output{food source with more nectar (higher fitness)}
    initialize $l$ food sources with position $\overrightarrow{X}_i(0)$ within boundaries $l_b$, $u_b$\;
    define (initial) best solution $\overrightarrow{G}$ as the food source with higher fitness\;
    \For{$n \in \{1,...,max\_it\}$}{
        \For{$i \in \{1,...,\ell\}$)}{
            define a new position for the $i$-th employed bee based on the current food sources\;
            \eIf{the new position has higher fitness than the current}{
                set as the new position for the employed bee\;
            }{
                increase abandonment counter\;
            }
        }
        \For{$i \in \{1,...,\ell\}$}{
            select a food source by roulette wheel selection based on the probability of such position to have more nectar in comparison with others\;
            define a position for the $i$-th onlooker bee based on the selected food source\;
            \eIf{the new position has higher fitness than the current}{
                set as the new position for the employed bee\;
            }{
                increase abandonment counter\;
            }
        }
        \If{abandonment counter exceeds $L$}{
            the exceeding food sources are abandoned and new (random) food sources are initialized by scout bees\;
            the abandonment counter is reset\;
        }
    }
    \caption{Artificial bee colony}
\label{alg:abc}  
\end{algorithm}

\section{}
\label{appendix:gwo}
\begin{algorithm}[H]
\SetKwInOut{Input}{input}\SetKwInOut{Output}{output}
  \Input{$f$, $l_b$, $u_b$, $\ell$, $max\_it$}
  \Output{position of the $\alpha$ wolf}
  initialize $\ell$ wolves with random position within boundaries $l_b$, $u_b$\;
  define the three higher fitness wolves as $\alpha$, $\beta$, and $\delta$\;
  \For{$n \in \{1,...,max\_it\}$}{
      \For{$i \in \{1,...,\ell \}$}{
        position is updated following the mean of the $\alpha$, $\beta$, and $\delta$ wolves\;
      }
      the three wolves with higher fitness are the new $\alpha$, $\beta$, and $\delta$\;
  }
  \caption{Grey wolf optimizer pseudo-code}
\label{alg:gwo}  
\end{algorithm}

\section{}
\label{appendix:igwo}
\begin{algorithm}[H]
\SetKwInOut{Input}{input}\SetKwInOut{Output}{output}
  \Input{$f$, $l_b$, $u_b$, $\ell$, $max\_it$}
  \Output{position of the $\alpha$ wolf}
  initialize $\ell$ wolves with random position within boundaries $l_b$, $u_b$\;
  define the three higher fitness wolves as $\alpha$, $\beta$, and $\delta$\;
  \For{$n \in \{1,...,max\_it\}$}{
      \For{$i \in \{1,...,\ell \}$}{
        the \textit{candidate} position is defined following the mean of the $\alpha$, $\beta$, and $\delta$ wolves (as in GWO)\;
        a neighborhood is constructed based on the current position and candidate position\;
        multi-neighbor learning is performed (DLH solution)\;
        choose best position between candidate and DLH for the update\;
      }
      the three wolves with higher fitness are the new $\alpha$, $\beta$, and $\delta$\;
  }
  \caption{Improved grey wolf optimizer pseudo-code}
\label{alg:igwo}  
\end{algorithm}

\newpage

 \bibliographystyle{elsarticle-num} 
 \bibliography{cas-refs}





\end{document}